\documentclass[letter]{aa}  
\usepackage{natbib}
\usepackage{orcidlink}
\usepackage{txfonts}
\usepackage{bm}
\bibpunct{(}{)}{;}{a}{}{,} 

\usepackage{graphicx}

\usepackage{hyperref}
\newcommand{\beq}{\begin{equation}}
\newcommand{\eeq}{\end{equation}}
\newcommand{\bea}{\begin{eqnarray}}
\newcommand{\eea}{\end{eqnarray}}

\newcommand{\cm}{\mathcal}
\newcommand{\mb}{\mathbf}
\newcommand{\al}{\langle}
\newcommand{\ar}{\rangle}

\definecolor{azul}{rgb}{0,0,.8}
\definecolor{rojo}{rgb}{1,0,0}
\definecolor{verde}{rgb}{0,.5,0}
\definecolor{violeta}{rgb}{.5,.0,1}
\definecolor{gris}{rgb}{.5,.5,.5}
\definecolor{marron}{rgb}{.4,.1,0}
\definecolor{naranja}{rgb}{1,.5,0}
\definecolor{bordo}{rgb}{.5,0,.2}
\definecolor{rojo}{rgb}{1,0,0}

\usepackage[normalem]{ulem}

\begin{document}

\title{Strong signature of right-handed circularly polarized photoionization close to the cyclotron line in the atmosphere of magnetic white dwarfs}

\titlerunning{Strong signature of right-circularly polarized absorption}

\author{Ren\'e D. Rohrmann\orcidlink{0000-0001-7209-3574}}

\institute{Instituto de Ciencias Astron\'omicas, de la Tierra y del
Espacio (CONICET-UNSJ), Av. Espa\~na 1512 (sur), 5400 San Juan, Argentina}
 
\abstract{
Magnetic fields break the symmetry of the interaction of atoms with photons with different polarizations, yielding chirality and anisotropy properties.
The dependence of the absorption spectrum on the polarization, a phenomenon known as dichroism, is present in the atmosphere of magnetic white dwarfs. Its evaluation for processes in the continuum spectrum has been elusive so far due to the absence of appropriate ionization equilibrium models and  incomplete data on photoionization cross sections. We combined rigorous solutions to the equilibrium of atomic populations with approximate cross sections to calculate the absolute opacity due to photoionization in a magnetized hydrogen gas. We predict a strong right-handed circularly polarized absorption ($\chi^+$) formed blueward of the cyclotron resonance for fields from about 14 to several hundred megagauss. In energies lower than the cyclotron fundamental, 
this absorption shows a deep trough with respect to linear and left-handed circular polarizations that steepens with the field strength. The jump in $\chi^+$ is due to the confluence of a large number of photoionization continua produced by right-handed circularly polarized transitions
from atomic states with a nonnegative magnetic quantum number toward different Landau levels.
} 


\maketitle
%

\section{Introduction}\label{s:intro}

About 800 magnetic white dwarfs (MWDs) have been discovered so far  \citep{ferrario2020}; most have hydrogen-rich atmospheres (similar to non-MWDs) and magnetic fields ($B$) from around $0.01$ up to nearly $10^3$ megagauss (MG). Analysis of a volume-limited sample suggests that 20\% of all white dwarfs have magnetic fields in their surfaces \citep{bagnulo2021}. Field intensities in MWDs are mainly detected via displacements of spectral lines and Zeeman splittings \citep{amorim2023}. 
They can also be detected via narrow- and broadband circular polarization measurements and spectropolarimetry; the latter is a valuable method for measuring fields when it is difficult to identify spectral lines \citep{berdyugin2022} or there is a complex line pattern (strong MWDs). 

Current model atmospheres fit the spectra of MWDs with a weak or moderate field strength, $B\la 50$~MG, relatively well but have trouble achieving satisfactory precision for stronger ones \citep{kulebi2009, hardy2023, vera2024}. Also, the description of the continuum polarization spectrum in highly MWDs remains incomplete. This is partly due to the lack of reliable opacity data for bound-free (bf) atomic transitions in strong magnetic fields. 
Detailed models of both flux and polarization spectra require magnetic-field-dependent photoionization opacities for atomic H.  Accurate photoionization cross sections have been obtained since the 1990s \citep{delande1991, wang1991, merani1995, zhao2007, zhao2021}. However, the available data are very limited and unsatisfactory for model atmosphere calculations, as they cover a small number of transitions over a sparse grid of field strengths and a usually incomplete polarization basis. Consequently, MWD models 
use an adapted form \citep{jordan1992} of the bf absorption first formulated by \citet{lamb1972} for weak fields ($B<10$ MG). 
This approach is justified for the linear Zeeman regime, where the field-perturbed, atomic  Hamiltonian is diagonal and so the matrix elements associated with bf radiative transitions are unaltered.
Such properties led to an ansatz to evaluating field-dependent cross sections in terms of their zero-field values, which are distributed among the various components according to the Wigner-Eckart theorem.
Curiously, to the best of our knowledge, a direct comparison between the cross sections resulting from this procedure and those from full quantum mechanical calculations has never been made. 
Moreover, no definitive conclusions about bf cross sections can be reached from spectrum fits unless they are self-consistently applied with field-dependent atomic populations. The absolute value of the bf opacity depends on the occupation numbers of many sublevels, which must be calculated in a thermodynamically consistent way for partially ionized, magnetized atmospheres. The lack of chemical equilibrium models for hydrogen gas in the MWD field regime has been a significant limitation of all model atmospheres published until recently. Nowadays, an equation of state fulfills this requirement \citep{vera2020}.

This Letter presents the first self-consistent, total photoionization absorption of a partially ionized hydrogen gas in the magnetic field regime of MWDs. A comparison of the cross sections of the updated Lamb-Sutherland ansatz with full quantum values is given for the first time (Sect. \ref{s:cross}). These calculations were combined with a detailed chemical model (Sect. \ref{s:popul}) to obtain the total photoionization opacity as a sum of all relevant transitions arising from the discrete state spectrum of magnetized hydrogen atoms for the three basic polarizations of light  (Sect. \ref{s:opac}). The prediction of a previously unknown dichroic feature is reported.

\section{Cross-section evaluation}\label{s:cross}

First, we consider zero-field conditions. The photoionization cross section in dipole and single-particle approximation is proportional to the energy of the absorbed photon, $\cm{E}$, and the matrix element of the dipole moment (in the usual notation):
\beq \label{basic}
\sigma^q_{nlm,kl'm'} = \text{cte} \;  \cm{E} \;
        |\al nlm |\mb{r}.\mb{e}_q| kl'm'\ar|^2,
\eeq
where $|nlm\ar$ is the bound initial state, $| kl'm'\ar$ the final state (the eigenstate for the ejected electron), $\mb{r}$  the electron position in the atom, and $\mb{e}_q$ the unit polarization vector of the incident photon. Transitions with $q =-1,0,+1$ are the so-called $\sigma^-$, $\pi$, and $\sigma^+$ transitions and are referred to as left-handed circular, linear, and right-handed circular polarizations, respectively. The main quantum number, $n,$ and its analogous for the continuum, $k,$ are related to the energies of the bound and continuum atomic states and to the energy of the absorbed photon, in Rydberg (Ry) units:
\beq \label{energy0}
\cm{E}_n=-n^{-2}, \hskip.2in \cm{E}_k=k^{-2}, \hskip.2in 
\cm{E} = \cm{E}_k-\cm{E}_n.
\eeq
According to the Wigner-Eckart theorem, the matrix element in Eq. (\ref{basic}) can be separated into geometrical and physical parts,
\beq \label{1a}
\sigma^q_{nlm,kl'm'} = \text{cte}\;\cm{E}\, 
   \cm{W}^{\;l,\;1,\; l'}_{m,\, q,\,m'} |\al nl\|\mb{r}\| kl'\ar|^2,
\eeq
where $\cm{W}$ denotes the square of the Wigner (3-$j$) coefficient,
\beq
\cm{W}^{\;l,\;1,\; l'}_{m,\, q,\,m'} \equiv
\left(\begin{array}{ccc} l& 1 & l' \\ m& q &-m' \end{array}\right)^2,
\eeq
and the rest of terms in Eq. (\ref{1a}) are independent of projections of angular momenta ($m$, $q,$ and $m'$).
The cross section for transitions $nl \rightarrow kl'$ follows the usual rule 
of summing  over the final states and averaging over the initial states,
\beq \label{1b}
\sigma_{nl,kl'}=\frac{1}{2l+1}\sum_{m m'} \sigma^q_{nlm,kl'm'}.
\eeq
Combining Eqs. (\ref{1a}) and (\ref{1b}), one obtains
\beq \label{sigwig}
\sigma^q_{nlm,kl'm'}=3(2l+1)
\cm{W}^{\;l,\;1,\; l'}_{m,\, q,\,m'} \sigma_{nl,kl'},
\eeq
where a sum property of the Wigner coefficients was used,
$\sum_{m,m'} \cm{W}^{\;l,\;1,\; l'}_{m,\, q,\,m'}=1/3$.
\begin{figure}
\includegraphics[width=.43\textwidth]{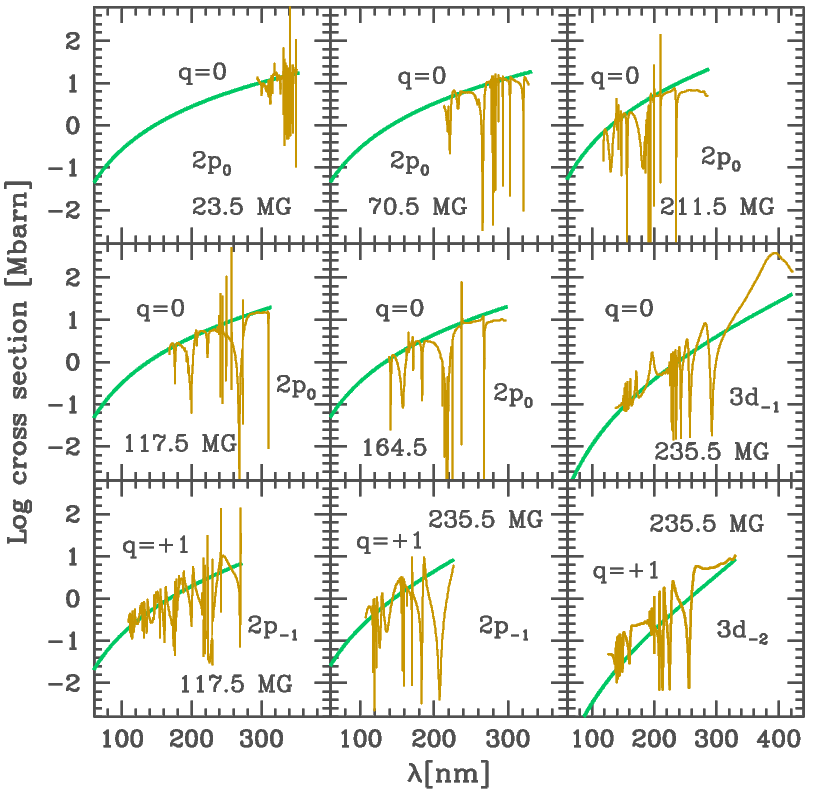}
\caption{Partial photoionization cross section in megabarns as a function of the light wavelength for different $nlm$ sublevels ($m_s=-\frac12$), polarizations ($q=0$ and $q=+1$), and magnetic field strengths (as indicated on the plot). Results from Eqs. (\ref{main}) and (\ref{lamb}) (thick green lines) are compared with solutions from \citet[thin orange lines]{zhao2007}.}
\label{f:zhao}
\end{figure}
On the other hand, $\sigma_{nl,kl'}$ can be written in terms of the mean cross section for transitions $n \rightarrow k$ ,
\beq\label{fnlk_v2}
\sigma_{nl,kl'}=\frac{Q_{nl,kl'}}{P_{nk}}\sigma_{n,k},\hskip.26in
\eeq
with $P_{nk}$ and $Q_{nl,kl'}$ polynomials on $k$ whose explicit expressions can be found in the literature \citep[e.g.,][]{menzel1935,hatanaka1946}.
With Eqs. (\ref{sigwig}) and (\ref{fnlk_v2}), summing over all final sublevels, we obtain the cross section for a given photon polarization ($q=m'-m$) and for allowed dipolar transitions ($l'=l\pm1$) from a specific bound sublevel to the continuum, 
\beq \label{main}
\sigma^q_{nlm,k} =\frac{3(2l+1)}{P_{nk}} 
 \left[\cm{W}^{\;l,\;1,\; l+1}_{m,\, q,\,m+q} Q_{nl,k(l+1)}
     + \cm{W}^{\;l,\;1,\; l-1}_{m,\, q,\,m+q}Q_{nl,k(l-1)}\right] 
  \sigma_{n,k}.
\eeq
Equation (\ref{main}) is exact in the absence of a magnetic field. The proposal of \citet{lamb1972} to evaluate the cross section when a magnetic field is present consists in assuming that the initial and final wave functions remain essentially unchanged in a small region of overlap such that the matrix element in Eq. (\ref{basic}) remains unaltered. The cross section can thus be approximated as
\beq \label{lamb}
\sigma(\cm{E})=\frac{\cm{E}}{\cm{E} -\Delta} \sigma_{0}(\cm{E} -\Delta),
\hskip.4in \Delta=\cm{E}_*-\cm{E}_0,
\eeq
where  $\cm{E}_*$ and $\cm{E}_0$ are the ionization thresholds with and without a magnetic field, and $\sigma_0$ denotes the field-free cross section. 
This approach is asymptotically correct at $B\rightarrow 0$ and, when combined with precise ionization threshold energies (Sect. \ref{s:opac}), provides a surprisingly good approximation at mean field strengths as high as 100-200 MG.

Figure \ref{f:zhao} compares results from Eqs. (\ref{main}) and  (\ref{lamb}) with precise evaluations from \citet{zhao2007} for the lowest field strengths published in that work.
The cross sections based on full quantum mechanic calculations show rich resonance structures that arise due to quasi-bound Coulombic states embedded in the Landau continua. The evaluations based on Eqs. (\ref{main}) and (\ref{lamb}) are able to reproduce the mean values of the full theory. 
The comparison is even better if one takes into account the fact that much of the resonance structure can play a minor role in the formation of spectra. Since the magnetic field on the stellar surface varies significantly, by up to a factor of two for a dipole field, most narrow resonances will be smeared out or smoothed as they are averaged over even a fraction of the visible disk. These results support the use of Eqs. (\ref{main}) and (\ref{lamb}), supplemented with accurate energy data, as an empirical approach that provides reasonable cross sections at moderately high field strengths beyond the linear Zeeman regime.

\section{Atomic energies and occupation numbers}\label{s:popul}

The Hamiltonian of a non-relativistic electron in a fixed Coulomb potential and a uniform magnetic field, $B$ (directed along the $z$-axis) can be written using cylindrical coordinates ($\phi,\rho,z$) and atomic units (except energy in Ry) as
\beq\label{HH}
H=H'+2\beta\left(\hat{l}_z+2\hat{s}_z\right),\hskip.3in 
 H'=-\Delta -\frac{2}{r}+\beta^2\rho^2,
\eeq
with $r=\sqrt{\rho^2+z^2}$, $\beta=B/B_0$ a dimensionless field strength ($B_0=4.70103\times 10^9$~G), and $\hat{l}_z$ and $\hat{s}_z$ the components of the orbital and spin angular momenta in the $z$-axis.  
Eigenstates of $H$ can be identified via the magnetic and spin quantum numbers ($m$ and $m_s=\pm\frac12$) complemented with the principal ($n$) and orbital ($l$) ones in the weak-field regime, and by the Landau number, $N,$ and the longitudinal quantum number, $\nu$ (associated with excitations of the wave function along the field lines) in the strong-field regime. The  correspondence between ($n,l$) and ($N,\nu$) is based on the non-crossing rule \citep{simola1978}, and their explicit relations are given in \citet{vera2020}. 

Eigenvalues of $H$ have contributions ($\cm{E}'$) from the reduced Hamiltonian, $H'$ (which contains the diamagnetic term, quadratic in $\beta$) and paramagnetic contributions (linear in $\beta$): 
\beq\label{eem}
\cm{E}=\cm{E}'+2\beta(m+2m_s).
\eeq
According to Eq. (\ref{eem}), states with $m>0$ lie $4\beta m$ above those with $-m$,
\beq \label{em}
\cm{E}_m=\cm{E}_{-m}+4\beta m.
\eeq
Hereafter, we use a compact notation, $\cm{E}_{m}\equiv \cm{E}_{nlmm_s}$, where the magnetic quantum number is highlighted because of its relevance in the discussion. High-precision values for $\cm{E}_{m\le 0}$  were provided by \citet{schimeczek2014b} and fitted in \citet{vera2020} for a large number of bound states as a function of $\beta$. On the other hand, continuum energies that define the ionization thresholds are specified by Landau states of a free electron with linear momentum, $p_z$, along the field 
\beq\label{ec}
\cm{E}_c = 4\beta \left(N+m_s+\frac12\right)+\frac{p_z^2}{2m_\text{e}}, \hskip.2in
N= n_\rho +\frac{|m'|+m'}2,
\eeq
with $n_\rho=0,1,...$, the radial quantum number of a Landau state.

\begin{figure}
\includegraphics[width=.44\textwidth]{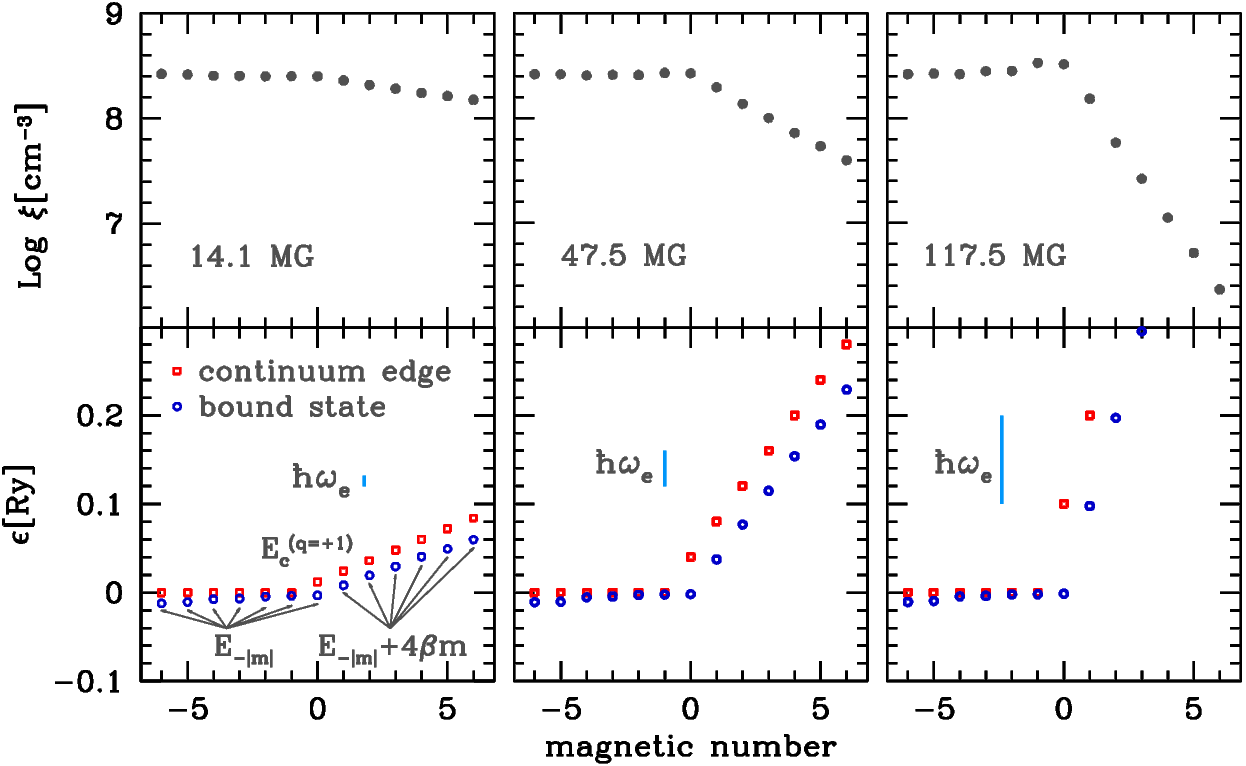}
\caption{Occupation numbers ($\xi$; top) and energies of bound energies (bottom) for sublevels $(n,l)=(11,6)$   for various $B$ (gas temperature $20000$~K and density $10^{-8}$~g/cm$^3$), shown as circles. The ionization edges (red squares) for $\sigma^+$ transitions and the size of the cyclotron energy are also shown in the lower panel.} 
\label{n11l6D}
\end{figure}   
The ionization equilibrium and atomic populations were calculated as described in \citet{vera2020}. Since the field strengths considered here are not too high, we used atomic internal partition in the form given by \citet{pavlov1993}, with finite-velocity effects incorporated using effective atomic masses. 
Most of the MWDs have fields between 10 and 100 MG, a range in which the onset of profound changes in the eigenenergy spectrum and the distribution of atomic populations takes place. As an illustrative case, Fig. \ref{n11l6D} shows occupation numbers and energies corresponding to sublevels $(n,l)=(11,6)$  for a temperature of $20000$~K. Energies of states with $m>0$ are increased by multiples of cyclotron energy $\hbar\omega_\text{e}=4\beta$ with respect to levels $\cm{E}_{-|m|}$ (Eq. (\ref{em})), and their populations decrease accordingly due to the usual Boltzmann factor. 
The abundance of atoms in spin-up states is 48\% at $B=14$~MG and falls to 30\% at $117$~MG. Around $32$~MG there is a peak of 3.5\% of atoms in metastable bound states ($m>0$) above the first Landau level.

\section{Absolute photoionization opacity}\label{s:opac}

\begin{figure}
\includegraphics[width=.45\textwidth]{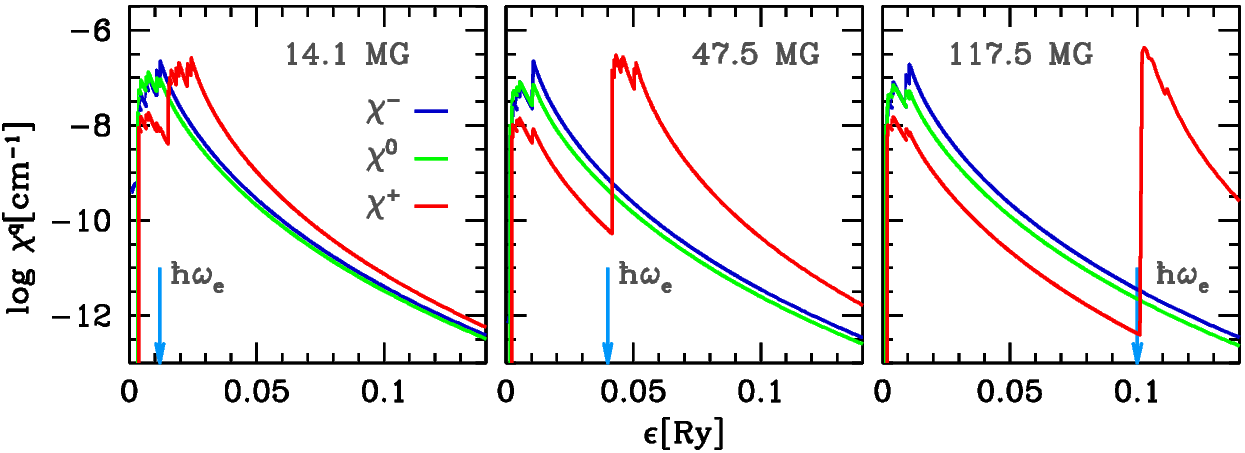} 
\caption{Contributions to the extinction $\chi^q$ from sublevels $(n,l)=(11,6)$ as a function of the photon energy, for the conditions shown in Fig. \ref{n11l6D}.} 
\label{n11l6S}
\end{figure}   
The monochromatic extinction coefficient due to photoionization for the three polarizations, $q=0,\pm1,$ is given by
\beq \label{chi}
\chi^q=\sum_{nlmm_s}\xi_{nlmm_s}\sigma^q_{nlmm_s,k},
\eeq
where $\xi_{nlmm_s}$ is the number density of particles in a bound state and the sum  extends over all bound states, which we label with the quantum numbers of an atom in the zero-field limit.
\begin{figure*}
\sidecaption
\includegraphics[width=12cm]{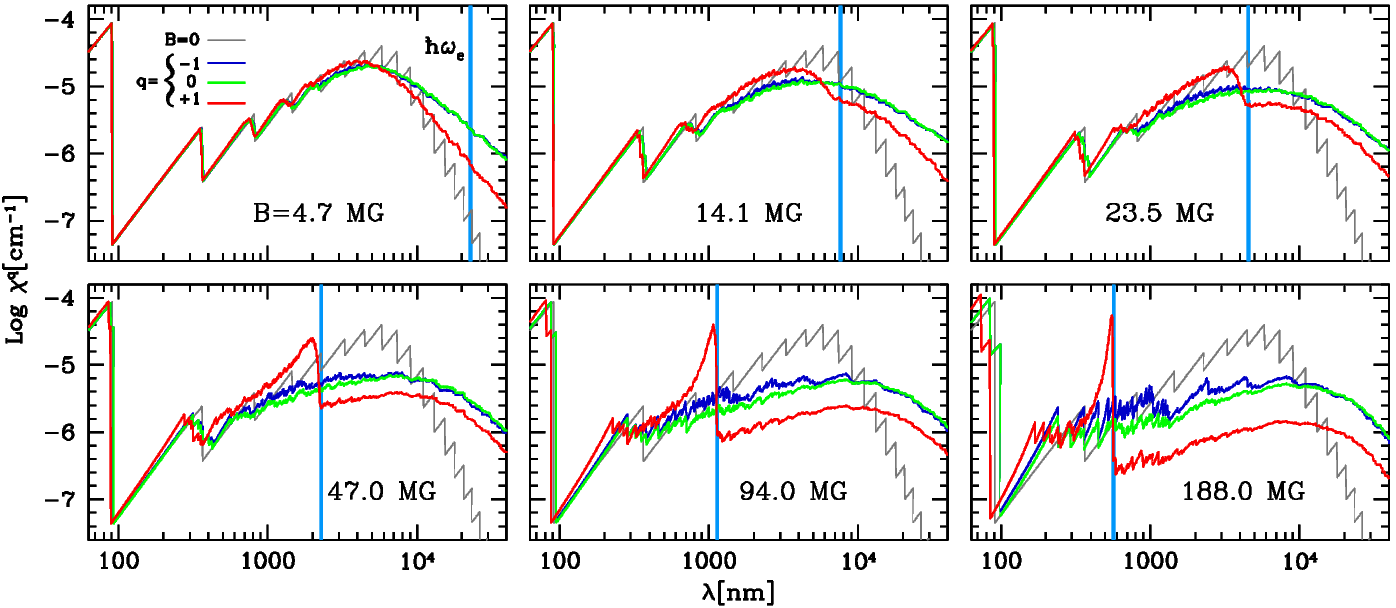}
\caption{Monochromatic extinction coefficient due to photoionization for the three basic polarizations, $q=0,\pm1$, and for a hydrogen gas with $T=20000$~K and $\rho=10^{-8}$~g~cm$^{-3}$. 
The extinctions for $B=0$ (thin gray lines) and cyclotron resonance (light blue line) are also shown.} 
\label{veff}
\end{figure*}   

First we consider transitions from spin-down levels.
The minimum photoionization energy, $\cm{E}_*$, of a bound state (given by Eqs. (\ref{em}) and (\ref{ec}) with $p_z=0$, $n_\rho=0$, $m_s=-\frac12$)
depends on the photon polarization ($q=m'-m$), which characterizes the corresponding continuum edge \citep{jordan1992,merani1995},
\beq\label{ebinding}
\cm{E}_* \equiv \cm{E}_c-\cm{E}_m= \begin{cases} 
-\cm{E}_{-|m|} + 4\beta, & q=+1~(m\ge 0), \\ 
\max\{0,-\cm{E}_{-|m|} - 4\beta\}, & q=-1~(m \ge 1), \\ 
-\cm{E}_{-|m|}, & \text{otherwise}. \\
\end{cases}
\eeq
In general, $\cm{E}_*$ equals the binding energy, $-\cm{E}_{-|m|}$, except in two notable cases. First, for right-handed circular polarization ($q=+1$) and $m\ge0,$ the continuum edge has an additional increment of $4\beta$, which is the cyclotron energy, $\hbar\omega_\text{e}$, expressed in Ry.
Second, for left-handed circular polarization ($q=-1$) and positive $m,$ the continuum threshold is reduced by $4\beta$, which can place the continuum edge near or even below the bound state for a strong enough field; in this case, photoionization takes place from zero photon energy.
This scheme is the same for spin-up states because the spin-field coupling affects both discrete and continuum states equally and the spin does not change in the transition;  the only difference is that the initial spin-down levels are statistically more populated and contribute more to the opacity. 

Figure \ref{n11l6S} shows the contributions to the opacity arising from sublevels at $(n,l)$=$(11,6)$ for different field strengths and polarizations.
At relatively low field strengths, $B<200$~MG ($\beta<0.05$), relevant binding energies, $-\cm{E}_{-|m|}$, are moderately modified around their zero-field values, $\text{Ry}/n^2$, which are lower than the cyclotron energy for $n>0.5/\sqrt{\beta}\ga 3$. This means that a great number of transitions from excited sublevels, especially  $\pi$ and $\sigma^-$ transitions, are shifted toward infrared and blue wavelengths as the field strength increases.  However, the most significant changes occur for $\sigma^+$ transitions from $B\approx 14$~MG; this is when the cyclotron energy becomes appreciably greater than most $-\cm{E}_{-|m|}$ values (as shown in Fig. \ref{n11l6D}). In such a case, $\sigma^+$ photoionization has two origins: transitions from states with negative $m$ have thresholds distributed as in the previous cases, while those from $m\ge 0$ move to energies higher by  $\hbar\omega_\text{e}=4\beta$.
As shown in Fig. \ref{n11l6S}, there is a  clear second jump in $\chi^+$ on the cyclotron fundamental, and it moves toward higher energies as the field size increases. This occurs with similar characteristics for sublevels originated from other $n$ and $l$, $n>(4\beta)^{-1/2}$; all those with $m\ge0$ reinforce the $\sigma^+$ continuum with an edge along the cyclotron line.

Figure \ref{veff} shows the total photoionization opacities at $q=0,\pm1$ produced by a hydrogen gas as a function of the photon wavelength ($\lambda$) and for various field strengths. They are compared with the zero-field opacity, which shows the usual discontinuities associated with initial levels $n$=1--17. At low strengths (4.7 MG), the magnetic field mainly affects highly excited states, so the largest opacity changes occur at infrared and longer wavelengths.
When $B$ exceeds 14~MG, a bulge in $\chi^+$ forms at $\lambda$ shorter than the cyclotron resonance. As the field strengthens, the bulge transforms into a well-defined jump followed by a deep trough at longer wavelengths.
$\chi^-$ opacities are very similar to  $\chi^0$  opacities, with minor differences arising from ionization rules and Wigner coefficients; both are larger than $\chi^+$ at wavelengths longer than the cyclotron line. The opposite occurs at shorter wavelengths, where the bulge of $\sigma^+$ transitions from $m\ge0$ sublevels takes place.

Figure \ref{veff} clearly shows how the photoionization continuum of a magnetized hydrogen gas becomes strongly dichroic ($q$-dependent), and the medium reveals handedness (circular chirality) through marked differences between $\chi^-$ and $\chi^+$. 
While the total abundance of neutral atoms (mostly in the ground state) is slightly affected by the magnetic field for the studied strengths (the Lyman continuum remains almost unchanged at $B\la100$~MG), the occupation numbers of the excited levels modeling the bf continuum in the visible and infrared spectra are strongly modified. Consequently, self-consistent calculations are needed to bring the continuum opacity into agreement with the chemical equilibrium for magnetized gases.  Such calculations reveal the formation of a strong, right-handed circularly polarized absorption next to the cyclotron line, formed by the superposition of several thousand continua coming from Hilbert subspaces $m\ge0$.
Much of the resonance pattern in the cross section of each isolated transition, as obtained in full quantum theory, is expected to be smoothed due to such a superposition. This provides additional support for the use of the ansatz based on Eqs. (\ref{main}) and (\ref{lamb}) for MWD atmospheres with low and moderate field strengths. 

The cyclotron resonance, which is completely right-handed circularly polarized, is displayed in Fig. \ref{veff} as a comparison. It was calculated following the prescription given by \citet{lamb1974}, which includes Doppler broadening and produces a very intense and very narrow absorption, as shown in the figure. Even with collision broadening included, this free-free opacity remains extremely sharp. However, \citet{martin1979} showed that the cyclotron absorption cannot produce very deep features in the spectrum for realistic field patterns on the stellar disk; it can at most produce a smooth, extended depression.
Similarly, the strong bf absorption reported here could also be smoothed by field spread.  While field broadening may reduce the effects of the proposed opacity in the flux spectra, the expected strong continuous circular polarization signature (with a strong asymmetry around $\hbar\omega_\text{e}$) may be detectable in polarization measurements.

The implications of these results on the flux and polarization spectra observed in MWDs will be investigated in a future study. 

\bibliographystyle{aa}
\bibliography{sigmaplus_v3}

\begin{thebibliography}{22}
\expandafter\ifx\csname natexlab\endcsname\relax\def\natexlab#1{#1}\fi

\bibitem[{{Amorim} {et~al.}(2023){Amorim}, {Kepler}, {K{\"u}lebi}, {Jordan}, \&
  {Romero}}]{amorim2023}
{Amorim}, L.~L., {Kepler}, S.~O., {K{\"u}lebi}, B., {Jordan}, S., \& {Romero},
  A.~D. 2023, \apj, 944, 56

\bibitem[{{Bagnulo} \& {Landstreet}(2021)}]{bagnulo2021}
{Bagnulo}, S. \& {Landstreet}, J.~D. 2021, \mnras, 507, 5902

\bibitem[{{Berdyugin} {et~al.}(2022){Berdyugin}, {Piirola}, {Bagnulo},
  {Landstreet}, \& {Berdyugina}}]{berdyugin2022}
{Berdyugin}, A.~V., {Piirola}, V., {Bagnulo}, S., {Landstreet}, J.~D., \&
  {Berdyugina}, S.~V. 2022, \aap, 657, A105

\bibitem[{{Delande} {et~al.}(1991){Delande}, {Bommier}, \& {Gay}}]{delande1991}
{Delande}, D., {Bommier}, A., \& {Gay}, J.~C. 1991, \prl, 66, 141

\bibitem[{{Ferrario} {et~al.}(2020){Ferrario}, {Wickramasinghe}, \&
  {Kawka}}]{ferrario2020}
{Ferrario}, L., {Wickramasinghe}, D., \& {Kawka}, A. 2020, Advances in Space
  Research, 66, 1025

\bibitem[{{Hardy} {et~al.}(2023){Hardy}, {Dufour}, \& {Jordan}}]{hardy2023}
{Hardy}, F., {Dufour}, P., \& {Jordan}, S. 2023, \mnras, 520, 6111

\bibitem[{{Hatanaka}(1946)}]{hatanaka1946}
{Hatanaka}, T. 1946, Japanese Journal of Astronomy and Geophysics, 21, 1

\bibitem[{{Jordan}(1992)}]{jordan1992}
{Jordan}, S. 1992, \aap, 265, 570

\bibitem[{{K{\"u}lebi} {et~al.}(2009){K{\"u}lebi}, {Jordan}, {Euchner},
  {G{\"a}nsicke}, \& {Hirsch}}]{kulebi2009}
{K{\"u}lebi}, B., {Jordan}, S., {Euchner}, F., {G{\"a}nsicke}, B.~T., \&
  {Hirsch}, H. 2009, \aap, 506, 1341

\bibitem[{{Lamb} \& {Sutherland}(1972)}]{lamb1972}
{Lamb}, F.~K. \& {Sutherland}, P.~G. 1972, in Line Formation in the Presence of
  Magnetic Fields, 183

\bibitem[{{Lamb} \& {Sutherland}(1974)}]{lamb1974}
{Lamb}, F.~K. \& {Sutherland}, P.~G. 1974, in Physics of Dense Matter, ed.
  C.~J. {Hansen}, Vol.~53, 265

\bibitem[{{Martin} \& {Wickramasinghe}(1979)}]{martin1979}
{Martin}, B. \& {Wickramasinghe}, D.~T. 1979, \mnras, 189, 69

\bibitem[{{Menzel} \& {Pekeris}(1935)}]{menzel1935}
{Menzel}, D.~H. \& {Pekeris}, C.~L. 1935, \mnras, 96, 77

\bibitem[{{Merani} {et~al.}(1995){Merani}, {Main}, \& {Wunner}}]{merani1995}
{Merani}, N., {Main}, J., \& {Wunner}, G. 1995, \aap, 298, 193

\bibitem[{{Pavlov} \& {Meszaros}(1993)}]{pavlov1993}
{Pavlov}, G.~G. \& {Meszaros}, P. 1993, \apj, 416, 752

\bibitem[{{Schimeczek} \& {Wunner}(2014)}]{schimeczek2014b}
{Schimeczek}, C. \& {Wunner}, G. 2014, \apjs, 212, 26

\bibitem[{{Simola} \& {Virtamo}(1978)}]{simola1978}
{Simola}, J. \& {Virtamo}, J. 1978, Journal of Physics B Atomic Molecular
  Physics, 11, 3309

\bibitem[{{Vera-Rueda} \& {Rohrmann}(2020)}]{vera2020}
{Vera-Rueda}, M. \& {Rohrmann}, R.~D. 2020, \aap, 635, A180

\bibitem[{{Vera-Rueda} \& {Rohrmann}(2024)}]{vera2024}
{Vera-Rueda}, M. \& {Rohrmann}, R.~D. 2024, \aap, 687, A141

\bibitem[{{Wang} \& {Greene}(1991)}]{wang1991}
{Wang}, Q. \& {Greene}, C.~H. 1991, \pra, 44, 7448

\bibitem[{{Zhao}(2021)}]{zhao2021}
{Zhao}, L.~B. 2021, \apjs, 254, 21

\bibitem[{{Zhao} \& {Stancil}(2007)}]{zhao2007}
{Zhao}, L.~B. \& {Stancil}, P.~C. 2007, \apj, 667, 1119

\end{thebibliography}

\end{document}